# 基於後量子密碼學的區塊鏈系統安全效能分析—以加密貨幣交換為例

Abel C. H. Chen

Information & Communications Security Laboratory,
Chunghwa Telecom Laboratories

摘要

　　現行的加密貨幣交換區塊鏈系統在錢包主要採用橢圓曲線密碼學金鑰對，而在交易主要採用橢圓曲線數位簽章演算法來產製簽章。因此，在量子電腦技術成熟時，現行的區塊鏈系統將面臨被量子計算攻擊的風險，並且量子電腦將可能仿冒橢圓曲線數位簽章演算法的簽章。有鑑於此，本研究將分析現行的區塊鏈系統存在抗量子計算弱點，並且提出基於後量子密碼學的區塊鏈系統，對現行的區塊鏈系統中存在的每個弱點進行改進和提升安全性。其中，本研究主要提出基於後量子密碼學錢包和基於後量子密碼學交易，可以運用後量子密碼學數位簽章演算法對交易輸入產製簽章，避免簽章被量子計算仿冒。在實驗結果顯示後量子密碼學數位簽章演算法 Dilithium 演算法的產製錢包、產製簽章、驗證簽章的效率高於現行區塊鏈系統的橢圓曲線數位簽章演算法，並且 Dilithium 演算法也具有較的安全等級。

關鍵詞：區塊鏈、後量子密碼學、效能分析

# The Security Performance Analysis of Blockchain System Based on Post-Quantum Cryptography — A Case Study of Cryptocurrency Exchanges

Abel C. H. Chen

Information & Communications Security Laboratory,
Chunghwa Telecom Laboratories

**Abstract**

　　The current blockchain system for cryptocurrency exchanges primarily employs elliptic curve cryptography (ECC) for generating key pairs in wallets, and elliptic curve digital signature algorithms (ECDSA) for generating signatures in transactions. Consequently, with the maturation of quantum computing technology, the current blockchain system faces the risk of quantum computing attacks. Quantum computers may potentially counterfeit signatures produced by ECDSA. Therefore, this study analyzes the vulnerabilities of the current blockchain system to quantum computing attacks and proposes a post-quantum cryptography (PQC)-based blockchain system to enhance security by addressing and improving each identified weakness. Furthermore, this study proposes PQC-based wallets and PQC-based transactions, utilizing PQC digital signature algorithms to generate PQC-based signatures for the inputs in PQC-based transactions, thereby preventing signatures from being counterfeited by quantum computing. Experimental results demonstrate that the efficiency of the Dilithium algorithm, a PQC digital signature algorithm, in producing wallets, generating signatures, and verifying signatures surpasses that of ECDSA in the current blockchain system. Furthermore, the Dilithium algorithm also exhibits a higher security level.

# 1.前言

近年來，隨著區塊鏈(Blockchain)系統和技術的普及，已經被廣泛應用到許多場域，並且國際電機電子工程師學會(Institute of Electrical and Electronics Engineers, IEEE)更在 2020 年後陸續制定多項標準，包含電子商務交易證據收集(IEEE 3802-2022 標準)(DFESC of IEEE CTS, 2022a)、電子合約(IEEE 3801-2022 標準)(DFESC of IEEE CTS, 2022b)、供應鏈金融(IEEE 2418.7-2021 標準)(BDLSC of IEEE CS, 2021)、數位資產管理(IEEE 2418.10-2022 標準)(BSC of IEEE CTS, 2022)、電子票據業務(IEEE 2142.1-2021 標準)(BSC of IEEE CTS, 2021)、碳交易應用(IEEE 3218-2022 標準)(BDLSC of IEEE CS, 2023)、加密貨幣交換(IEEE 2140.1-2020 標準)(BSC of IEEE CTS, 2020)等。

雖然區塊鏈系統可以提供分散式的架構，並且通過密碼學技術保障交易的不可否認性、完整性等，但大部分區塊鏈系統裡的數位簽章演算法仍是架構在橢圓曲線密碼學的基礎上(El-Zawawy, Brighente, & Conti, 2023; Zhou et al., 2023; Velliangiri et al., 2023)。然而，隨著量子電腦技術的成熟，橢圓曲線密碼學將面臨被破解的風險(Shor, 1997)，所以具有量子計算能力的攻擊者將可以破解出橢圓曲線密碼學私鑰，並且用橢圓曲線密碼學私鑰仿冒合法簽章。因此，開始有部分區塊鏈系統開始採用後量子密碼學方法來提升安全性，包含基於格(Lattice-based)密碼學的區塊鏈系統(Saha et al., 2023)、基於雜湊(Hash-based)密碼學的區塊鏈系統(Yi, 2022)。除此之外，美國國家標準與技術研究院(National Institute of Standards and Technology, NIST)在 2016 年開始徵求後量子密碼學(Post-Quantum Cryptography, PQC)，並制定了數個後量子密碼學演算法，包含有金鑰封裝機制(Key Encapsulation Mechanism, KEM)演算法 Kyber、數位簽章演算法(Digital Signature Algorithm, DSA) Dilithium、Falcon、以及 SPHINCS+ (Alagic et al., 2022)。

有鑑於提升區塊鏈系統在抗量子計算攻擊成的能力，本研究將分析現行的區塊鏈系統裡使用的密碼學方法，並分析抗量子計算攻擊的能力。除此之外，本研究將把目前已經通過美國國家標準與技術研究院認定為標準的後量子密碼學方法結合到區塊鏈系統，並對系統效能進行比較和分析。本研究貢獻主要包含有：

1. 設計和實作基於後量子密碼學的區塊鏈系統。

2. 驗證和比較各個後量子密碼學標準在區塊鏈系統中的效能。

3. 驗證和比較多個工作量證明(Proof of Work, PoW)參數在區塊鏈系統中的效能。

基於後量子密碼學的區塊鏈系統安全效能分析—以加密貨幣交換為例

本文分為五個章節。第二節將討論現行的區塊鏈系統，以及區塊鏈系統中的資料結構。第三節提出基於後量子密碼學的區塊鏈系統，並且從資料結構和案例說明。第四節將對本研究提出的方案進行實驗比較，分別從產製錢包效率、產製簽章效率比較、驗證簽章效率比較、以及工作量證明安全性分析等面向討論。第五節論述本研究的貢獻和探討未來研究方向。

## 2.區塊鏈系統

本研究主要參考 Kuznetsov 實作的區塊鏈系統(Kuznetsov, 2021)，並且在其基本上改進。在 2.1 節中將先說明現行的區塊鏈系統，描述錢包、交易、區塊、以及區塊鏈的資料結構和採用的密碼學方法，以及在 2.2 節討論現行的工作量證明方法及安全性。

## 2.1 區塊鏈系統資料結構

本節將介紹錢包(Wallet)、交易(Transaction)、區塊(Block)、以及區塊鏈的資料結構及運作原理。

### 2.1.1 錢包

在產製錢包時，主要將產生一對基於橢圓曲線密碼學(Elliptic Curve Cryptography, ECC)的金鑰對，包含基於橢圓曲線密碼學私鑰(ECC-based Private Key)和基於橢圓曲線密碼學公鑰(ECC-based Public Key)，如圖 1 所示。後續在交易時，將可以用基於橢圓曲線密碼學私鑰產製對交易內容的簽章；當其他使用者想確認交易是否成立且屬實時，可以用基於橢圓曲線密碼學公鑰驗證簽章。其中，將以基於橢圓曲線密碼學公鑰的雜湊(Hash)值作為錢包地址(Address)。

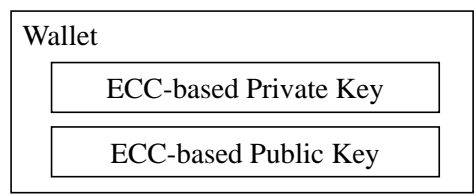

**圖 1　錢包**

### 2.1.2 交易

交易的內容主要包含交易識別碼(Transaction ID)、交易輸入(The Inputs of Transaction)、交易輸出(The Outputs of Transaction)、以及交易的產製時間(Generation Time)，如圖 2 所示。

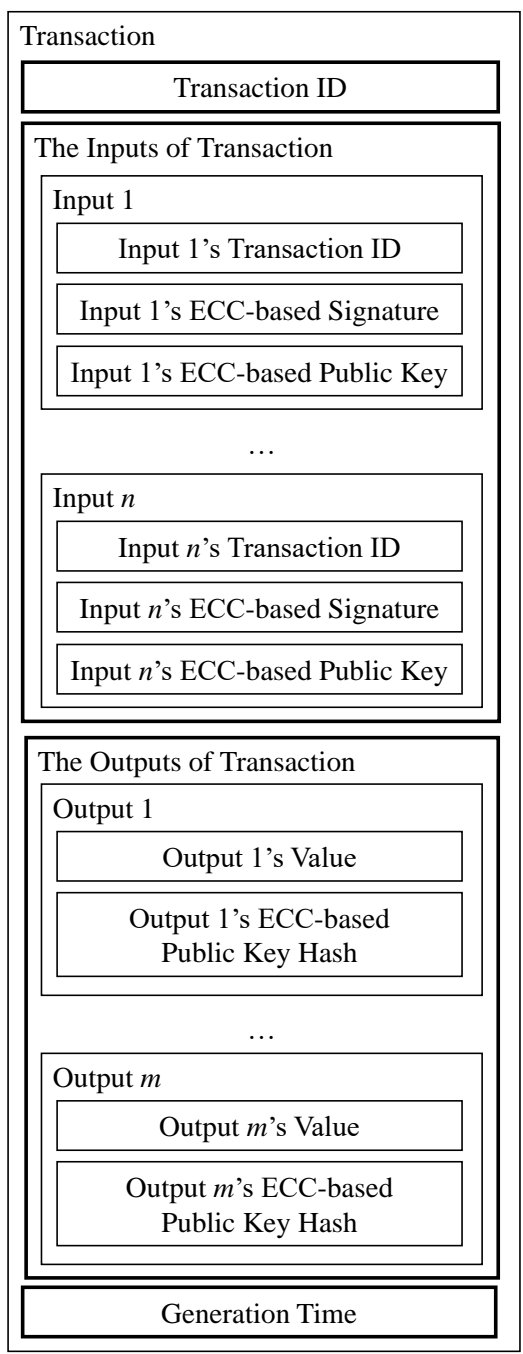

**圖 2　交易**

　　交易輸入是指加密貨幣擁有者的加密貨幣來源，而加密貨幣來源是來自加密貨幣擁有者之前的交易所獲取，並且有可能來自過去的多筆交易獲取。因此，在交易輸入中包含多筆輸入(Input)，並且加密貨幣擁有者移轉加密貨幣給另一位使用者，所以加密貨幣擁有者要用其基於橢圓曲線密碼學私鑰對每筆輸入產製基於橢圓曲線密碼學簽章(ECC-based Signature)。每筆輸入將包含交易識別碼、基於橢圓曲線密碼學簽章、基於橢圓曲線密碼學公鑰。

　　交易輸出是指加密貨幣擁有者移轉加密貨幣給另一位使用者後，雙方加密貨

幣的狀態。例如，交易輸出包含交易前的加密貨幣擁有者和交易後的加密貨幣擁持者等，所以有多筆輸出。每筆輸出將包含金額(Value)及對應貨幣擁持者的基於橢圓曲線密碼學公鑰雜湊值(也就是錢包地址)。

### 2.1.3 區塊

在區塊鏈系統中，進行交易時，需要把交易放到區塊中和產生區塊鏈。在本節(2.1.3 節)介紹區塊的資料結構，並在下一節(2.1.4 節)介紹區塊鏈。區塊包含區塊雜湊(Block Hash)、前一個區塊的區塊雜湊(Previous Block Hash)、交易集(Transactions)、區塊的產製時間、以及偽隨機數(Nonce)，如圖 3 所示。

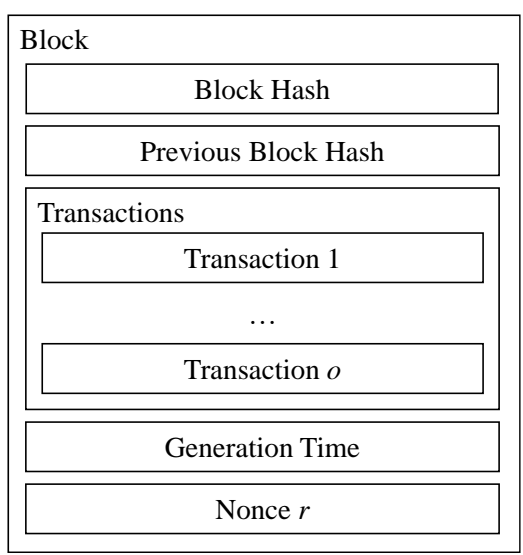

**圖 3　區塊**

當發生一筆交易時，需把交易加入到一個新區塊的"交易集"欄位，並且設置區塊的產製時間，在這個新區塊加入"前一個區塊的區塊雜湊"欄位以形成區塊鏈。設定好上述欄位值後,為提供完整性驗證,會對上述的區塊內容計算雜湊值。其中，為提升安全性會進行工作量證明，設置難度目標長度 $L$，對區塊內容計算偽隨機數 $r$ 次後的雜湊值作為區塊雜湊。工作量證明的詳細計算將於 2.2 節中說明。

### 2.1.4 區塊鏈

在區塊鏈系統中，每個區塊都存放區塊雜湊、前一個區塊的區塊雜湊，通過指向前一個區塊的的區塊雜湊來形成區塊鏈。因此，以第 $i$ 個區塊為例，在"區塊雜湊"欄位存放第 $i$ 個區塊的雜湊值，"前一個區塊的區塊雜湊"欄位存放第 $i$-1 個區塊的雜湊值，如圖 4 所示。

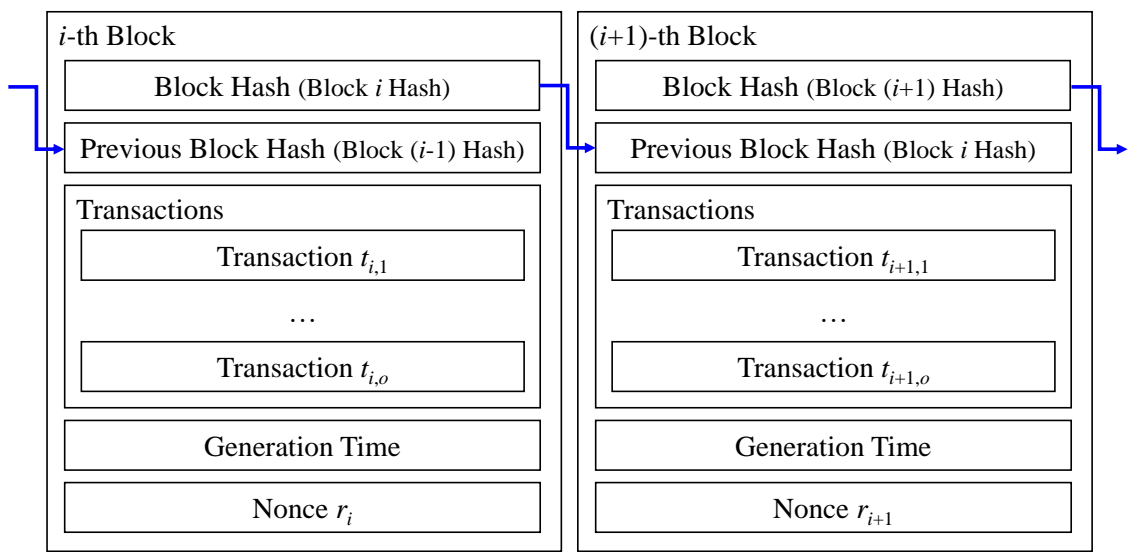

**圖 4 區塊鏈**

當使用者想驗證區塊鏈上的任一區塊完整性時，可以通過取得區塊內容，並對區塊內容計算偽隨機數 $r$ 次後的雜湊值與"區塊雜湊"欄位值比對，如果一致表示內容沒有被竄改過。因此，通過區塊鏈可以確認鏈上的每個區塊的完整性，並從而確定區塊中的交易屬實。

## 2.2 工作量證明

為建立區塊完整性，主要建構在工作量證明的計算基礎上，也就是挖礦，如圖 5 所示。其中，假設區塊內容為 $b$、雜湊函數為 $Hash(b)$、執行 $r$ 次雜湊函數表示為 $Hash^r(b)$、難度目標長度 $L$。首先取得區塊內容 $b$，然後計算區塊內容 h 的雜湊值 $Hash(b)$，判斷雜湊值的前 $L$ 個位元(bit)是否皆為 0；如果雜湊值的前 $L$ 個位元沒有全部皆為 0，則對雜湊值代入雜湊再做一次雜湊值，直到前 $L$ 個位元全部皆為 0 才停止。停止時，根據執行的雜湊函數次數作為 $r$ 值，也就是 $Hash^r(b)$ 的值符合前 $L$ 個位元全部皆為 0，並且把 $r$ 值設定到"偽隨機數"欄位。

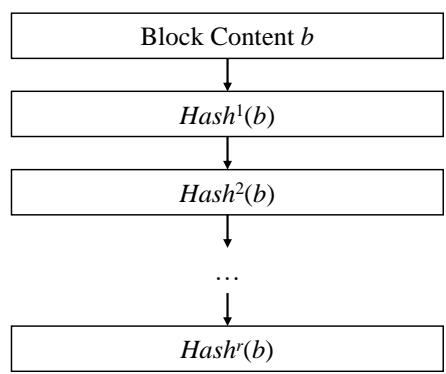

**圖 5 工作量證明**

## 2.3 安全效能討論

本節分別從現行的區塊鏈系統錢包、交易、區塊及其工作量證明來討論抗量子計算攻擊的能力，分述如下。

(1). 錢包：由於現行的區塊鏈系統錢包主要採用基於橢圓曲線密碼學私鑰和基於橢圓曲線密碼學公鑰，所以將有可能被量子計算攻擊有破解。因此，現行的區塊鏈系統錢包<u>存在抗量子計算弱點(**Quantum-Vulnerable**)</u>。

(2). 交易：由於現行的區塊鏈系統交易主要採用基於橢圓曲線密碼學簽章，但由於將有可能被量子計算攻擊有破解，所以將有可能被仿冒基於橢圓曲線密碼學簽章。因此，現行的區塊鏈系統交易<u>存在抗量子計算弱點</u>。

(3). 區塊、區塊鏈、工作量證明：由於現行的區塊鏈系統中的區塊及其工作量證明是建構在雜湊函數(如：安全雜湊演算法-256(Secure Hash Algorithms-256, SHA-256))的基礎上，而目前量子計算攻擊尚無法破解安全雜湊演算法(Alagic et al., 2022)，所以具備抗量子計算的能力。因此，工作量證明主要需要考量的是難度目標長度 $L$，當難度目標長度 $L$ 太短，則安全性較低，容易被攻擊成功；當難度目標長度 $L$ 太長，則計算時間(即挖礦時間)很久，需耗費較多計算資源。

## 3. 基於後量子密碼學的區塊鏈系統

為提升區塊鏈系統抵抗量子計算攻擊的能力，本研究提出基於後量子密碼學的區塊鏈系統，針對 2.3 節中提及存在抗量子計算弱點的部分加入後量子密碼學方法來提升安全性。在 3.1 節中描述本研究提出的區塊鏈系統資料結構，並且在 3.2 節提供案例說明。

## 3.1 本研究提出的區塊鏈系統資料結構

本節主要對錢包和交易提供後量子密碼學方案，分述於 3.1.1 節和 3.1.2 節。在 3.1.3 節中描述區塊、區塊鏈、工作量證明的安全性探討和改進方法。

### 3.1.1 錢包

本研究提出基於後量子密碼學錢包主要包含基於後量子密碼學私鑰(PQC-based Private Key)和基於後量子密碼學公鑰(PQC-based Public Key)，如圖 6 所示。在產製基於後量子密碼學錢包時，產生一對基於後量子密碼學的金鑰對，包含基於後量子密碼學私鑰和基於後量子密碼學公鑰。後續在交易時，將可以用

基於後量子密碼學私鑰產製對交易內容的簽章；當其他使用者想確認交易是否成立且屬實時，可以用基於後量子密碼學公鑰驗證簽章。其中，將以基於後量子密碼學公鑰的雜湊值作為錢包地址。

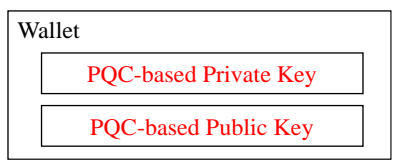

**圖 6　本研究提出基於後量子密碼學錢包**

## 3.1.2 交易

圖 7 為本研究提出基於後量子密碼學交易的資料結構，主要包含基於後量子密碼學簽章(PQC-based Signature)和基於後量子密碼學公鑰，詳述如下。

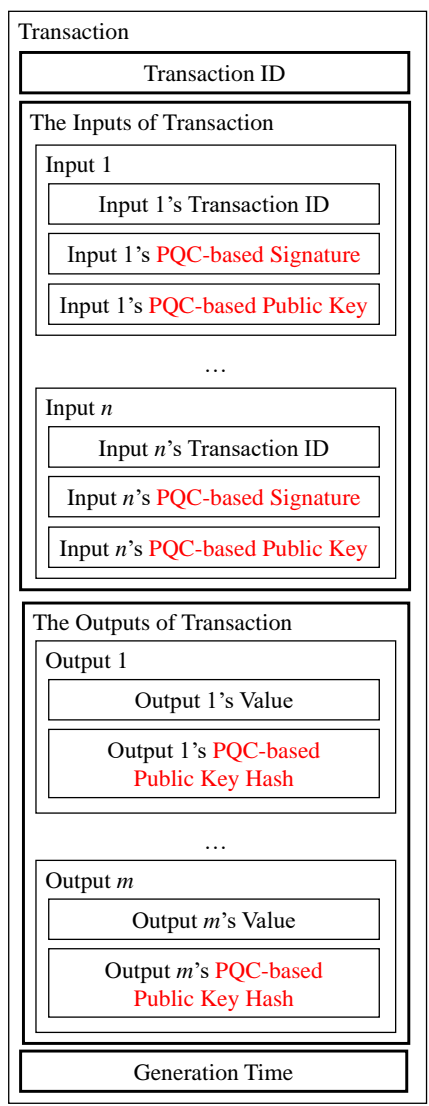

**圖 7　本研究提出基於後量子密碼學交易**

為避免被量子計算仿冒簽章，所以交易所使用的簽章需是基於後量子密碼學簽章，通過基於後量子密碼學錢包中的基於後量子密碼學私鑰對每一筆交易輸入產製基於後量子密碼學簽章，並且把基於後量子密碼學公鑰一併存放到交易輸入中。而在交易輸出中，由於基於後量子密碼學錢包的錢包地址已經是基於後量子密碼學公鑰的雜湊值，所以做對應的修改。

### 3.1.3 區塊、區塊鏈、工作量證明

由於現行的區塊鏈系統中的區塊、區塊鏈、工作量證明具備抗量子計算的能力，所以目前尚不需要修改密碼學演算法。然而，難度目標長度 $L$ 將會是需要衡量的主要因子，所以本研究將在第 4 節對不同的難度目標長度進行效率比較。

除此之外，有鑑於工作量證明主在計算前 $L$ 個位元全部皆為 0 時的雜湊值，所以本研究亦探索"前 $L$ 個位元全部皆為 0"、"前 $L$ 個位元全部皆為 1"的機率。雖然雜湊函數產生出來的雜湊值，每一個位元出現 0 或 1 的機率應該要接近均勻分佈。然而，在實際的計算上，每一個位元出現 0 或 1 的機率不完全是均勻分佈，所以本研究將在 4.5 探索安全的工作量證明演算法。

## 3.2 案例說明

本節提供案例說明本研究提出的基於後量子密碼學的區塊鏈系統，展示基於後量子密碼學錢包、基於後量子密碼學交易、以及區塊和區塊鏈的運作流程。

### 3.2.1 情境說明

在案例中，假設有 4 位使用者(分別為 Alice、Bob、Cara、以及 Davis)分別進行加密貨幣交換。在案例中的初始狀態假設和交易情境假設如下，並且在 3.2.2 節中說明交易情境所產生的基於後量子密碼學交易和區塊及區塊鏈的關聯。

初始狀態假設：

(1). Alice 在交易 Transaction ID $t_a$ 已獲取加密貨幣 5 元，且 Alice 僅有 5 元。

(2). Cara 在交易 Transaction ID $t_c$ 已獲取加密貨幣 7 元，且 Cara 僅有 7 元。

(3). Bob 和 David 加密貨幣皆為 0 元。

交易情境假設：

(1). Transaction ID $t_j$：Alice 移轉加密貨幣 5 元給 Bob。

(2). Transaction ID $t_{j+1}$：Cara 移轉加密貨幣 7 元給 Bob。

(3). Transaction ID $t_{j+2}$：Bob 移轉加密貨幣 12 元給 David。

### 3.2.2 區塊鏈系統資料結構說明

根據 3.2.1 節描述的交易情境，將依序產生圖 8 的基於後量子密碼學交易和圖 9 的基於後量子密碼學區塊鏈，詳述如下。

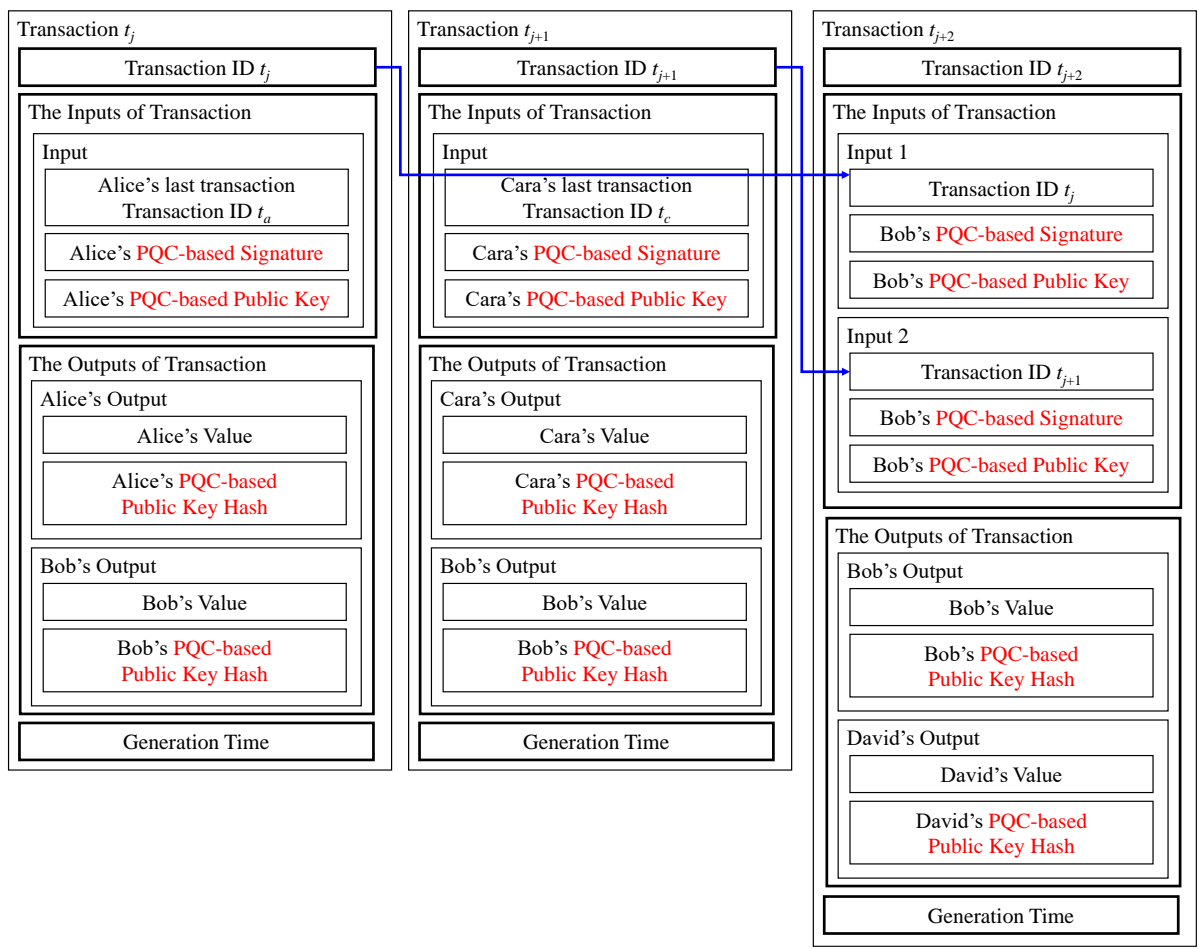

圖 8　案例中基於後量子密碼學交易

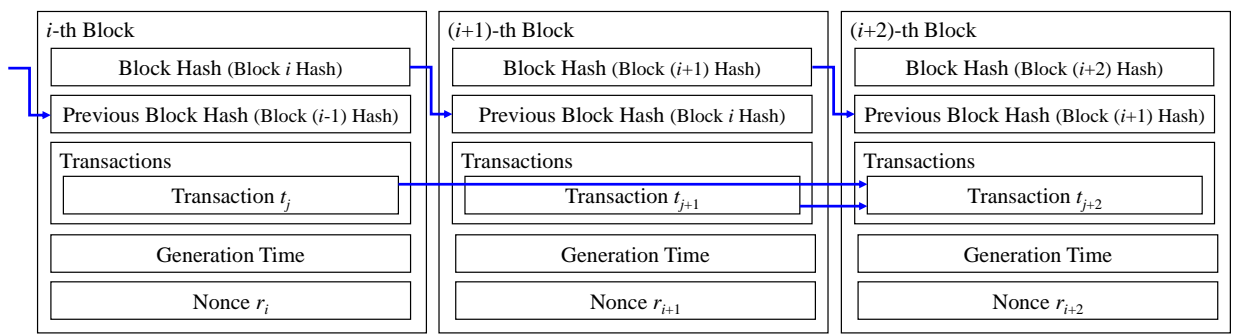

圖 9　案例中基於後量子密碼學區塊鏈

案例中交易情境所產生的基於後量子密碼學交易說明如下。

(1). Transaction ID $t_j$ (Alice 移轉加密貨幣 5 元給 Bob)

由於 Alice 在交易 Transaction ID $t_a$ 已獲取加密貨幣 5 元，所以當 Alice 要移轉加密貨幣 5 元給 Bob，可以用交易 Transaction ID $t_a$ 獲取的加密貨幣 5 元來移轉給 Bob。因此，產製交易 Transaction $t_j$，在交易輸入的欄位中設定一筆輸入，在這筆輸入的"交易識別碼"欄位是 Transaction ID $t_a$，並且在"基於後量子密碼學簽章"欄位存放用 Alice 基於後量子密碼學私鑰對這筆輸入的簽章，證實確實由 Alice 所進行的交易；以及在"基於後量子密碼學公鑰"欄位存放 Alice 基於後量子密碼學公鑰，作為後續驗證簽章使用。

在輸出的部分主要影響 Alice 的錢包和 Bob 的錢包，所以交易輸出包含有 2 筆輸出。在第 1 筆輸出的"基於後量子密碼學公鑰雜湊"欄位是 Alice 的錢包地址，"金額"欄位是 0，代表 Alice 在這筆交易後錢包的金額是 0 元。在第 2 筆輸出的"基於後量子密碼學公鑰雜湊"欄位是 Bob 的錢包地址，"金額"欄位是 5，代表 Bob 在這筆交易後錢包獲取的金額是 5 元。

(2). Transaction ID $t_{j+1}$ (Cara 移轉加密貨幣 7 元給 Bob)

由於 Cara 在交易 Transaction ID $t_c$ 已獲取加密貨幣 7 元，所以當 Cara 要移轉加密貨幣 7 元給 Bob，可以用交易 Transaction ID $t_c$ 獲取的加密貨幣 7 元來移轉給 Bob。因此，產製交易 Transaction $t_{j+1}$，在交易輸入的欄位中設定一筆輸入，在這筆輸入的"交易識別碼"欄位是 Transaction ID $t_c$，並且在"基於後量子密碼學簽章"欄位存放用 Cara 基於後量子密碼學私鑰對這筆輸入的簽章，證實確實由 Cara 所進行的交易；以及在"基於後量子密碼學公鑰"欄位存放 Cara 基於後量子密碼學公鑰，作為後續驗證簽章使用。

在輸出的部分主要影響 Cara 的錢包和 Bob 的錢包，所以交易輸出包含有 2 筆輸出。在第 1 筆輸出的"基於後量子密碼學公鑰雜湊"欄位是 Cara 的錢包地址，"金額"欄位是 0，代表 Cara 在這筆交易後錢包的金額是 0 元。在第 2 筆輸出的"基於後量子密碼學公鑰雜湊"欄位是 Bob 的錢包地址，"金額"欄位是 7，代表 Bob 在這筆交易後錢包獲取的金額是 7 元。

(3). Transaction ID $t_{j+2}$ (Bob 移轉加密貨幣 12 元給 David)

由於 Bob 移轉加密貨幣 12 元給 David，並且 Bob 沒有單一筆交易獲取 12 元以上的紀錄。因此，Bob 需結合多筆交易(即 Transaction $t_j$ 和 Transaction $t_{j+1}$ 兩筆結合)才有 12 元可以移轉給 Bob。產製交易 Transaction $t_{j+2}$，並且交易 Transaction $t_{j+2}$ 的交易輸入有多筆輸入。在第 1 筆輸入的"交易識別碼"欄位是 Transaction ID $t_j$，並且在"基於後量子密碼學簽章"欄位存放用 Bob 基於後量子密碼學私鑰對這筆輸入的簽章，證實確實由 Bob 所進行的交易；以及在"基於後量子密碼學公鑰"欄位存放 Bob 基於後量子密碼學公鑰，作為後續驗證簽章使用。在第 2 筆輸入的"交易識別碼"欄位是 Transaction ID $t_{j+1}$，並且與第 1 筆輸入相似，用 Bob 基於後量子密碼學私鑰產製簽章，並各別存放基於後量子密碼學簽章和基於後量子密碼學公鑰。

在輸出的部分主要影響 Bob 的錢包和 David 的錢包，所以交易輸出包含有 2 筆輸出。在第 1 筆輸出的"基於後量子密碼學公鑰雜湊"欄位是 Bob 的錢包地址，"金額"欄位是 0，代表 Bob 在這筆交易後錢包的金額是 0 元。在第 2 筆輸出的"基於後量子密碼學公鑰雜湊"欄位是 David 的錢包地址，"金額"欄位是 12，代表 David 在這筆交易後錢包獲取的金額是 12 元。

案例中交易情境所產生的基於後量子密碼學區塊鏈說明如下。

(1). 第 $i$ 個區塊(Alice 移轉加密貨幣 5 元給 Bob)

當發生交易 Transaction $t_j$ 時，產製第 $i$ 個區塊用以驗證交易 Transaction $t_j$ 的完整性。在第 $i$ 個區塊的"前一個區塊的區塊雜湊"欄位存放第 $i$-1 個區塊的區塊雜湊，指向前一個區塊，形成區塊鏈裡的區塊；在"交易集"欄位存放交易 Transaction $t_j$，並且在"產製時間"欄位設置產製區塊時的時間。之後根據工作量證明演算法以區塊內容產製區塊雜湊和偽隨機數 $r_i$。

(2). 第 $i$+1 個區塊(Cara 移轉加密貨幣 7 元給 Bob)

當發生交易 Transaction $t_{j+1}$ 時，產製第 $i$+1 個區塊用以驗證交易 Transaction $t_{j+1}$ 的完整性。與前述作法相似，在第 $i$+1 個區塊的"前一個區塊的區塊雜湊"欄位存放第 $i$ 個區塊的區塊雜湊、"交易集"欄位存放交易 Transaction $t_{j+1}$、"產製時間"欄位存放產製時的時間；並且在"區塊雜湊"欄位和"偽隨機數"欄位根據工作量證明演算法分別產製適合的值後存放。

(3). 第 $i$+2 個區塊(Bob 移轉加密貨幣 12 元給 David)

當發生交易 Transaction $t_{j+2}$ 時，產製第 $i$+2 個區塊用以驗證交易 Transaction $t_{j+2}$ 的完整性。與前述作法相似，在第 $i$+2 個區塊的"前一個區塊的區塊雜湊"欄位存放第 $i$+1 個區塊的區塊雜湊、"交易集"欄位存放交易 Transaction $t_{j+2}$、"產製時

間"欄位存放產製時的時間；並且在"區塊雜湊"欄位和"偽隨機數"欄位根據工作量證明演算法分別產製適合的值後存放。

# 4.系統驗證和實驗結果討論

本節對本研究提出的基於後量子密碼學的區塊鏈系統進行安全效能驗證。4.1 節先介紹實驗環境，4.2 節、4.3 節、4.4 節各別比較產製錢包效率、產製簽章效率、以及驗證簽章效率。並且在 4.5 節對工作量證明進行相關參數的安全性分析。最後，4.6 節小結實驗結果和發現。

## 4.1 實驗環境

在實驗環境中，本研究在 Kuznetsov 實作的區塊鏈系統(Kuznetsov, 2021)基礎上改寫成 Java 程式語言，並在一台 Windows 10 企業版的筆記型電腦開發本研究提出的基於後量子密碼學的區塊鏈系統。其中，筆記型電腦的軟硬體詳細規格是 CPU Intel(R) Core(TM) i7-10510U、記憶體 16 GB、OpenJDK 18.0.2.1、以及函式庫 Bouncy Castle Release 1.72。

本研究主要比較現行的區塊鏈系統使用橢圓曲線數位簽章演算法(Elliptic Curve Digital Signature Algorithm, ECDSA)(Baliker et al., 2024)，以及比較已經被美國國家標準與技術研究院(National Institute of Standards and Technology, NIST)列為標準的後量子密碼學數位簽章演算法，包含有 Dilithium、Falcon、SPHINCS+ (Alagic et al., 2022)。

為了比較在不同安全等級(Security Level)下各種演算法計算效率，本研究採用美國國家標準與技術研究院所規範的安全等級標準(NIST, 2023)，共可分為 5 個等級。其中，安全等級主要用來衡量後量子密碼學演算法抗量子計算攻擊的水平，與傳統的安全強度(Security Strength)(Barker, 2020)不同。因此，RSA 密碼學演算法和橢圓曲線密碼學演算法等不具抗量子計算攻擊的密碼學演算法在安全等級劃分都直接標記為"不安全"。

## 4.2 產製錢包效率比較

在產製錢包的過程中主要將產製密碼學金鑰對，即產製一組私鑰和一組對應的公鑰。各個演算法及其對應的參數產製錢包的計算時間如表 1 所示。其中，由實驗結果可知，雖然現行區塊鏈系統採用的橢圓曲線數位簽章演算法可以在 1.51 毫秒~1.84 毫秒產製一個錢包，但不具備抗量子計算攻擊的能力，所以安全等級

為"不安全"。Dilithium 演算法分別有安全等級 2、3、5 的參數組合(即 dilithium2_aes、dilithium3_aes、以及 dilithium5_aes)，並且可以觀察到可以在 0.61 毫秒~1.52 毫秒產製一個錢包，相於較現行區塊鏈系統的橢圓曲線數位簽章演算法有更高的安全等級且更短的產製錢包時間。本研究亦對 Falcon 演算法(包含參數組合 falcon_512、falcon_1024)和 SPHINCS+演算法(包含參數組合 shake_128f、shake_192f、shake_256f)進行比較；雖然 Falcon 演算法和 SPHINCS+演算法可以提供抗量子計算攻擊的能力，但產製錢包時間顯著高於橢圓曲線數位簽章演算法和 Dilithium 演算法。

表 1　產製錢包效率比較(單位：毫秒)

| 演算法 | 參數 | 安全等級 | 產製錢包時間 |
|---|---|---|---|
| ECDSA (Baliker et al., 2024) | P-256 | 不安全 | 1.51 |
| ECDSA (Baliker et al., 2024) | P-384 | 不安全 | 1.71 |
| ECDSA (Baliker et al., 2024) | P-521 | 不安全 | 1.84 |
| Dilithium | dilithium2_aes | 2 | **0.61** |
| Dilithium | dilithium3_aes | 3 | 0.81 |
| Dilithium | dilithium5_aes | 5 | 1.52 |
| Falcon | falcon_512 | 1 | 12.55 |
| Falcon | falcon_1024 | 5 | 32.58 |
| SPHINCS+ | shake_128f | 1 | 5.32 |
| SPHINCS+ | shake_192f | 3 | 7.53 |
| SPHINCS+ | shake_256f | 5 | 19.4 |

## 4.3 產製簽章效率比較

在產製交易的過程中主要將需要用加密貨幣擁有者的私鑰對交易輸入產製簽章。各個演算法及其對應的參數產製簽章的計算時間如表 2 所示。其中，由實驗結果可知，雖然現行區塊鏈系統採用的橢圓曲線數位簽章演算法產製簽章的時間介於 1.24 毫秒~1.88 毫秒，但安全等級為"不安全"。由於每個後量子密碼學演算法都具備抗量子計算攻擊的能力，為公平比較每個後量子密碼學，可以觀察在安全等級 5 時的表現。實驗結果顯示 Dilithium 在參數組合 dilithium5_aes 的產製簽章時間是 2.66 毫秒、Falcon 在參數組合 falcon_1024 的產製簽章時間是 3.22 毫秒、SPHINCS+在參數組合 shake_256f 的產製簽章時間是 386.98 毫秒。因此，由此可知在相同安全等級的情況下，Dilithium 演算法可以提供最快的產製簽章效率。除此之外，可以觀察到 Dilithium 在參數組合 dilithium2_aes 的產製簽章時間

是 1.14 毫秒比橢圓曲線數位簽章演算法(參數組合 P-256)、Falcon 演算法(參數組合 falcon_512)的產製簽章時間短；也就是 Dilithium 演算法(參數組合 dilithium2_aes)可以提供比橢圓曲線數位簽章演算法(參數組合 P-256)、Falcon 演算法(參數組合 falcon_512)更快且更安全的效能。

表 2 產製簽章效率比較(單位：毫秒)

| 演算法 | 參數 | 安全等級 | 產製簽章時間 |
| --- | --- | --- | --- |
| ECDSA (Baliker et al., 2024) | P-256 | 不安全 | 1.24 |
| ECDSA (Baliker et al., 2024) | P-384 | 不安全 | 1.52 |
| ECDSA (Baliker et al., 2024) | P-521 | 不安全 | 1.88 |
| Dilithium | dilithium2_aes | 2 | **1.14** |
| Dilithium | dilithium3_aes | 3 | 1.65 |
| Dilithium | dilithium5_aes | 5 | 2.66 |
| Falcon | falcon_512 | 1 | 1.94 |
| Falcon | falcon_1024 | 5 | 3.22 |
| SPHINCS+ | shake_128f | 1 | 121.61 |
| SPHINCS+ | shake_192f | 3 | 193.49 |
| SPHINCS+ | shake_256f | 5 | 386.98 |

## 4.4 驗證簽章效率比較

在驗證交易的過程中主要將需要用每筆交易輸入中所附的公鑰對該筆交易輸入的簽章進行驗證。各個演算法及其對應的參數驗證簽章的計算時間如表 3 所示。由實驗結果可以觀察到 Falcon 演算法的驗證簽章時間最短，介於 0.28 毫秒~0.46 毫秒；而 Dilithium 演算法的驗證簽章時間次短，介於 0.46 毫秒~1.3 毫秒。Falcon 演算法和 Dilithium 演算法都能提供比現行區塊鏈系統採用的橢圓曲線數位簽章演算法(驗證簽章時間介於2.82毫秒~3.5毫秒)更安全且更短的驗證簽章時間。

為公平比較每個後量子密碼學演算法的驗證簽章時間，可以觀察在安全等級 5 時的表現。實驗結果顯示 Dilithium 在參數組合 dilithium5_aes 的驗證簽章時間是 1.3 毫秒、Falcon 在參數組合 falcon_1024 的驗證簽章時間是 0.46 毫秒、SPHINCS+在參數組合 shake_256f 的驗證簽章時間是 10.59 毫秒。因此，由此可知在相同安全等級的情況下，Falcon 演算法可以提供最快的產製簽章效率，所以 Falcon 演算法可以更適用於需要高頻率驗證簽章的區塊鏈系統。

表 3　驗證簽章效率比較(單位：毫秒)

| 演算法 | 參數 | 安全等級 | 驗證簽章時間 |
|---|---|---|---|
| ECDSA (Baliker et al., 2024) | P-256 | 不安全 | 2.82 |
| ECDSA (Baliker et al., 2024) | P-384 | 不安全 | 3.39 |
| ECDSA (Baliker et al., 2024) | P-521 | 不安全 | 3.5 |
| Dilithium | dilithium2_aes | 2 | 0.46 |
| Dilithium | dilithium3_aes | 3 | 0.64 |
| Dilithium | dilithium5_aes | 5 | 1.3 |
| Falcon | falcon_512 | 1 | **0.28** |
| Falcon | falcon_1024 | 5 | 0.46 |
| SPHINCS+ | shake_128f | 1 | 7.3 |
| SPHINCS+ | shake_192f | 3 | 10.62 |
| SPHINCS+ | shake_256f | 5 | 10.59 |

## 4.5 工作量證明安全性分析

由於工作量證明主要設定難度目標長度 $L$，對區塊內容重覆迭代到雜湊函數計算，直到前 $L$ 個位元全部皆為 0 才停止。如果雜湊函數產製出來的雜湊值服從均勻分佈的情況下，則"前 $L$ 個位元全部皆為 0"機率和"前 $L$ 個位元全部皆為 1"機率沒有顯著差異，並接近 $1/2^L$。因此，本研究對工作量證明安全性分析主要分為兩個實驗：(1).不同難度目標長度的平均工作量證明計算時間、(2)雜湊值前 $L$ 個位元是否服從均勻分佈，實驗結果分述如下。

### 4.5.1 實驗(1)：不同難度目標長度的平均工作量證明計算時間

為驗證在不同度目標長度時的工作量證明計算時間，分別考慮難度目標長度 $L$ 為 4、8、12、16、20、以及 24，各執行 100 次交易的平均工作量證明計算時間。由實驗結果可以觀察可知，隨著難度目標長度的倍數增加時，平均工作量證明計算時間是呈指數級增加。其中，難度目標長度 $L$ 為 4 時，平均工作量證明計算時間僅需 0.31 毫秒，可以很快計算出結果；然而，這也代表攻擊者比較容易且快速的攻擊成功。而當難度目標長度 $L$ 為 24 時，平均工作量證明計算時間則上升到 4196.17 毫秒，不論是對傳統電腦或量子電腦都會隨著難度目標長度的增加，而有指數級的計算時間增加。有鑑於此，在部署區塊鏈系統時，應該採用較長的難度目標長度來提升安全性。

表 4  不同難度目標長度的平均工作量證明計算時間比較(單位：毫秒)

| 難度目標長度 | 工作量證明計算時間 |
| --- | --- |
| 4 | 0.31 |
| 8 | 1.06 |
| 12 | 2.41 |
| 16 | 17.22 |
| 20 | 321.4 |
| 24 | 4196.17 |

### 4.5.2 實驗(2)：雜湊值前 $L$ 個位元是否服從均勻分佈

為驗證雜湊值前 $L$ 個位元是否服從均勻分佈，本研究以難度目標長度 $L$ 為 8，並且執行 10000 次交易後觀察雜湊值前 $L$ 個位元來進行統計檢定。在此統計檢定中，如果把雜湊值以 16 進位制來表示，雜湊值前 2 個 16 進位制值為 00、01、…、FF，共 256 種組合。在均勻分佈的情況下，每一種組合發生的機率是 $1/2^8 = 1/256 = 0.391\%$；與真實發生的每一種組合相比計算後得到卡方值為 0.0011，P-value 為 1，所以在統計上每一種組合發生的機率與期望值沒有顯著差異，近似均勻分佈。

雖然在統計檢定上沒有顯著差異，但本研究仍根據難度目標長度 $L$ 為 4 和 8 時，觀察"雜湊值前 $L$ 個位元全部皆為 0"機率和"雜湊值前 $L$ 個位元全部皆為 1"的機率，如表 5 所示。由實驗數據顯示，"雜湊值前 $L$ 個位元全部皆為 1"機率較"雜湊值前 $L$ 個位元全部皆為 0"機率略低一些，所以如果在工作量證明計算時判斷"雜湊值前 $L$ 個位元全部皆為 1"的難度會更高且安全性會更高。

表 5  雜湊值前 $L$ 個位元(單位：毫秒)

| 難度目標長度 $L$ | 雜湊值前 $L$ 個位元皆為 0 | 雜湊值前 $L$ 個位元皆為 1 |
| --- | --- | --- |
| 4 | 6.249% (均勻分佈期望值：6.25%) | 6.241% (均勻分佈期望值：6.25%) |
| 8 | 0.391% (均勻分佈期望值：0.391%) | 0.390% (均勻分佈期望值：0.391%) |

## 4.6 小結與討論

總結本研究的實驗發現，以作為後續部署基於後量子密碼學的區塊鏈系統的實作參考。

(1). 在後量子密碼學數位簽章演算法的選擇上可以考慮採用 Dilithium 演算法。Dilithium 演算法在產製錢包、產製簽章、驗證簽章上的效率都高於現行區塊鏈系統的橢圓曲線數位簽章演算法。並且，Dilithium 演算法具備抗量子計算攻擊的能力，所以 Dilithium 演算法的安全等級也優於現行區塊鏈系統的橢圓曲線數位簽章演算法。

(2). 後量子密碼學數位簽章演算法 Falcon 演算法在驗證簽章的效率上是最快的。因此，如果有需要大量驗證簽章的區塊鏈系統使用情境時，可以採用 Falcon 演算法。

(3). 根據本研究實驗發現"雜湊值前 $L$ 個位元全部皆為 1"機率較"雜湊值前 $L$ 個位元全部皆為 0"機率略低一些。因此，在工作量證明的計算上建議可以修改為判斷"雜湊值前 $L$ 個位元全部皆為 1"來提升安全性。

## 5.結論與未來研究

本研究分析現行的區塊鏈系統存在抗量子計算弱點，並且提出基於後量子密碼學的區塊鏈系統，對現行的區塊鏈系統中存在的每個弱點進行改進和提升安全性。其中，本研究主要提出基於後量子密碼學錢包，在基於後量子密碼學錢包中建立基於後量子密碼學金鑰對；此外，本研究亦提出基於後量子密碼學交易，在基於後量子密碼學交易中運用後量子密碼學數位簽章演算法對交易輸入產製簽章，避免簽章被量子計算仿冒。最後，本研究探索工作量證明的安全性，並分析雜湊值發生的機率，以及在 4.6 節提出建議和改進方式。

在未來研究可以將本研究提出的基於後量子密碼學的區塊鏈系統實際部署到商用的加密貨幣交換應用上，以提升現行商用系統的安全性。

## 參考文獻

Alagic, G. et al. (2022). Status Report on the Third Round of the NIST Post-Quantum Cryptography Standardization Process. *National Institute of Standards and Technology, NIST IR 8413-upd1*. DOI: 10.6028/NIST.IR.8413-upd1

基於後量子密碼學的區塊鏈系統安全效能分析─以加密貨幣交換為例

Baliker, C., Baza, M., Alourani, A., Alshehri, A., Alshahrani, H., Choo, K. -K. R. (2024). On the Applications of Blockchain in FinTech: Advancements and Opportunities. *IEEE Transactions on Engineering Management*, to be published, DOI: 10.1109/TEM.2022.3231057.

Barker, E. (2020). *Recommendation for key management: part 1 – general*. National Institute of Standards and Technology, NIST SP 800-57 Part 1 Rev. 5. DOI: 10.6028/NIST.SP.800-57pt1r5

Blockchain and Distributed Ledgers Standards Committee (BDLSC) of IEEE Computer Society (CS). (2021). IEEE Standard for the Use of Blockchain in Supply Chain Finance. *IEEE Std 2418.7-2021*, 1-25, DOI: 10.1109/IEEESTD.2021.9599622.

BDLSC of IEEE CS. (2023). IEEE Standard for Using Blockchain for Carbon Trading Applications. *IEEE Std 3218-2022*, 1-31, DOI: 10.1109/IEEESTD.2023.10057186.

Blockchain Standards Committee (BSC) of IEEE CTS. (2022). IEEE Standard for Blockchain based Digital Asset Management. *IEEE Std 2418.10-2022*, 1-19, DOI: 10.1109/IEEESTD.2022.9810177.

BSC of IEEE CTS. (2021). IEEE Recommended Practice for E-Invoice Business Using Blockchain Technology. *IEEE Std 2142.1-2021*, 1-18, DOI: 10.1109/IEEESTD.2021.9381780.

BSC of IEEE CTS. (2020). IEEE Standard for General Requirements for Cryptocurrency Exchanges. *IEEE Std 2140.1-2020*, 1-18, DOI: 10.1109/IEEESTD.2020.9248667.

Digital Finance and Economy Standards Committee (DFESC) of IEEE Consumer Technology Society (CTS). (2022a). IEEE Standard for Application Technical Specification of Blockchain-based E-Commerce Transaction Evidence Collecting. *IEEE Std 3802-2022*, 1-24, DOI: 10.1109/IEEESTD.2022.9745865.

DFESC of IEEE CTS. (2022b). Standard for Blockchain-based Electronic Contracts. *IEEE Std 3801-2022*, 1-26, DOI: 10.1109/IEEESTD.2022.9745868.

El-Zawawy, M. A., Brighente, A., Conti, M. (2023). Authenticating Drone-Assisted Internet of Vehicles Using Elliptic Curve Cryptography and Blockchain. *IEEE Transactions on Network and Service Management*, 20(2), 1775-1789, DOI: 10.1109/TNSM.2022.3217320.

Kuznetsov, I. (Updated on June 26, 2021). Blockchain in Go, GitHub. Retrieved on January 12, 2024. URL: https://github.com/Jeiwan/blockchain_go.

NIST. (Updated on December 19, 2023). Security (Evaluation Criteria), Post-Quantum Cryptography, NIST. Retrieved on January 12, 2024. URL: https://csrc.nist.gov/projects/post-quantum-cryptography/post-quantum-cryptography-standardization/evaluation-criteria/security-(evaluation-criteria)

Saha, R. et al. (2023). A Blockchain Framework in Post-Quantum Decentralization. *IEEE Transactions on Services Computing*, 16(1), 1-12, DOI: 10.1109/TSC.2021.3116896.

Shor, P. W. (1997). Polynomial-time Algorithms for Prime Factorization and Discrete Logarithms on a Quantum Computer. *SIAM Journal on Computing*, 26(5), 1484-1509. DOI: 10.1137/S0097539795293172

Velliangiri, S. et al. (2022). An Efficient Lightweight Privacy-Preserving Mechanism for Industry 4.0 Based on Elliptic Curve Cryptography. *IEEE Transactions on Industrial Informatics*, 18(9), 6494-6502, DOI: 10.1109/TII.2021.3139609.

Yi, H. (2022). Secure Social Internet of Things Based on Post-Quantum Blockchain. *IEEE Transactions on Network Science and Engineering*, 9(3), 950-957, DOI: 10.1109/TNSE.2021.3095192.

Zhou, Z., Tian, Y., Xiong, J., Ma, J. Peng, C. (2023). "Blockchain-Enabled Secure and Trusted Federated Data Sharing in IIoT. *IEEE Transactions on Industrial Informatics*, 19(5), 6669-6681, DOI: 10.1109/TII.2022.3215192.